\documentclass{emulateapj}
\usepackage{graphicx}
\begin{document}

\def\etal{et al.\ \rm}
\def\aGI{\alpha_{\rm gt}}
\def\ba{\begin{eqnarray}}
\def\ea{\end{eqnarray}}
\def\etal{et al.\ \rm}

\voffset=-1.9cm

\title{Viscosity prescription for gravitationally unstable 
accretion disks}

\author{Roman R. Rafikov\altaffilmark{1}}
\altaffiltext{1}{Department of Astrophysical Sciences,
Princeton University, Ivy Lane, Princeton, NJ 08540; 
rrr@astro.princeton.edu}

%%%%%%%%%%%%%%%%%%%%%%%%%%%%%%%%%%%%%%%%%%%%%%%%%%%%%%%%%%%

\begin{abstract}
Gravitationally unstable accretion disks emerge in a variety 
of astrophysical contexts --- giant planet formation, 
FU Orioni outbursts, feeding of AGNs, and the origin of Pop 
III stars. When a gravitationally unstable disk is unable 
to cool rapidly it settles into a quasi-stationary, 
fluctuating gravitoturbulent state, in which its Toomre $Q$ 
remains close to a constant value $Q_0\sim 1$. Here we develop 
an analytical formalism describing the evolution of such a 
disk, which is based on the assumptions of $Q=Q_0$ and local 
thermal equilibrium. Our approach works in the presence of 
additional sources of angular momentum transport (e.g. MRI), 
as well as external irradiation. Thermal balance 
dictates a unique value of the gravitoturbulent stress 
$\alpha_{\rm gt}$ driving disk evolution, which is a 
function of the local surface density 
and angular frequency. We compare this approach with other 
commonly used gravitoturbulent viscosity prescriptions,
which specify the explicit dependence of stress $\alpha_{\rm gt}$ 
on Toomre $Q$ in an ad hoc fashion, and identify the ones 
that provide consistent results. We nevertheless argue that 
our $Q=Q_0$ approach is more flexible, robust, and 
straightforward, and should be given preference in 
applications. We illustrate this with a couple of analytical 
calculations --- locations of the snow line and of the outer 
edge of the dead zone in a gravitoturbulent protoplanetary 
disk --- which clearly show the simplicity and versatility 
of the $Q=Q_0$ approach.  
\end{abstract}

\keywords{accretion, accretion disks --- instabilities --- 
(stars:) planetary systems: protoplanetary disks --- 
(galaxies:) quasars: general}

%%%%%%%%%%%%%%%%%%%%%%%%%%%%%%%%%%%%%%%%%%%%%%%%%%%%%%%%%%%
\section{Introduction.}  
\label{sect:intro}
%%%%%%%%%%%%%%%%%%%%%%%%%%%%%%%%%%%%%%%%%%%%%%%%%%%%%%%%%%%

Astrophysical accretion disks can be prone to gravitational 
instability (hereafter GI) when their temperature is low 
and surface density is high (Safronov 1960; Toomre 1964). 
In particular, GI is conceivable in the outer parts 
of protoplanetary disks (Cameron 1978; Boss 1998) at 
the early stages, when the disks are still massive because 
of ongoing infall. Accretion 
outbursts of FU Orioni stars can be driven by the 
gravito-magnetic cycle in the dense, gravitationally unstable 
parts of the protoplanetary disk (Audard \etal 2014). Outer 
parts of AGN disks accreting at high $\dot M$ are 
also expected to become gravitationally 
unstable far from the black hole (Paczynski 1978a, 1978b; 
Kozlowski \etal 1979; 
Kolykhalov \& Sunyaev 1980; Goodman 2003). Turk \etal (2009), 
Clark \etal (2011), Greif \etal (2012) find gravitationally 
unstable disks around young Population III stars in their 
simulations of star formation at redshifts of $z\sim 20-30$. 

Possibility of the GI in a gaseous disk is characterized 
by the so-called Toomre $Q$ parameter defined as (Toomre 1964)
\ba
Q\equiv \frac{\Omega c_s}{\pi G\Sigma},
\label{eq:Q}
\ea
where $\Sigma$ and $c_s=(k_{\rm B}T/\mu)^{1/2}$ are the surface 
density and sound speed 
of the disk, and $\Omega=(GM/r^3)^{1/2}$ is the angular frequency 
(assuming Keplerian rotation around a central object with mass $M$). 
In the linear regime GI sets in as $Q\to Q_0$ from above, where 
$Q_0\approx 1-1.5$ is the threshold value suggested by numerical 
experiments (Boss 2002; Cossins \etal 2009, 2010). 

Operation of the GI is accompanied by enhanced angular momentum 
transport in the disk driven by the non-axisymmetric gravitational 
torques. As a result, GI results in mass redistribution in
the disk, driving its evolution. Energy dissipation caused by 
the application of gravitational stresses heats up the disk
and tends to oppose the GI. 

Because of this feedback, the nonlinear outcome of the GI 
sensitively depends on another dimensionless parameter --- the product 
of the local cooling time in the disk $t_{\rm cool}$ and $\Omega$. 
Gammie (2001) demonstrated that when cooling is fast and 
$\Omega t_{\rm cool}\lesssim \beta_{\rm crit}\sim 1$, the 
disk disintegrates into a number of bound self-gravitating 
structures. Such fragmentation has been invoked by Boss (1998) 
to explain the origin of giant planets (see Rafikov (2005, 2007) 
for constraints on this scenario).
Goodman \& Tan (2004) and Levin (2007) suggested that fragmentation 
of gravitationally unstable quasar disks should result in formation 
of massive stars migrating through the disk. Clark \etal (2011), 
Greif \etal (2012), Latif \& Schleicher (2014) propose that 
fragmentation 
of protostellar disks around Pop III stars can produce low mass 
extremely metal poor stars that can survive until present days.

In the opposite limit of long cooling time 
$\Omega t_{\rm cool}\gtrsim \beta_{\rm crit}$ 
it was shown by Gammie (2001) that the disk settles into a 
quasi-stationary state of {\it gravitoturbulence}. In this regime 
surface density fluctuates in time, but the disk state 
averaged over the period longer than the dynamical timescale 
$\Omega^{-1}$ remains the same. 
The time-averaged value of the Toomre Q parameter in a
gravitoturbulent disk is around $Q_0$, i.e. the disk maintains 
itself in a marginally stable state. This general picture has been 
confirmed with global simulations by different groups (Rice \etal 
2003, 2005; Durisen \etal 2007; Cossins \etal 2009,2010).

The critical value of the cooling time $\beta_{\rm crit}\Omega^{-1}$ 
corresponding to the transition between the gravitoturbulent 
and fragmenting regimes, depends on a variety of factors. One of 
the key determinants is
the equation of state (EOS) of the gas, with softer EOS promoting 
fragmentation and resulting in higher $\beta_{\rm crit}$
(Rice \etal 2005; Jiang \& Goodman 2011). Opacity transitions,
e.g. due to dust grain evaporation, also affect
critical $t_{\rm cool}$ (Johnson \& Gammie 2003), as well as 
other forms of the temperature dependence of opacity 
(Cossins \etal 2010). External irradiation (Rice \etal 2011) 
and the details of the disk structure (Meru \& Bate 2011a) 
may also affects the value of $\beta_{\rm crit}$. 

Some concerns have 
been raised regarding the convergence of gravitoturbulent disk 
simulations and the existence of the well-defined 
$\beta_{\rm crit}$, as its value has been claimed (Meru \& Bate 
2011b; Paardekooper 2012) to vary with the grid resolution. 
However, later this non-convergence has 
been traced mainly to the numerical effects (Paardekooper 2011; 
Meru \& Bate 2012; Rice \etal 2014).  

It has also been debated whether fragmentation ensues because 
rapid cooling facilitates collapse of unstable fragments, or 
because the disk can withstand only certain amount of stress and
fragments when the dimensionless stress parameter $\alpha$ (Shakura \& 
Sunyaev 1973) exceeds a threshold value $\alpha_{\rm crit}$. 
Since in thermal equilibrium in the absence of external 
energy inputs (Gammie 2001)
\ba
\alpha=\frac{4}{9\gamma(\gamma-1)}
\left(\Omega t_{\rm cool}\right)^{-1},
\label{eq:rel}
\ea
where $\gamma$ is the adiabatic index of gas, the values of 
$\alpha_{\rm crit}$ and $\beta_{\rm crit}$ are directly 
connected. Comparing simulations with different $\gamma$ 
(different EOS) Rice \etal (2005) noted that 
$\alpha_{\rm crit}$ is essentially independent of $\gamma$ 
and is about 0.06. This led them to suggest that the primary 
reason for fragmentation is the maximum stress that can be 
sustained by the disk. However, simulations with external 
irradiation (Rice \etal 2011) suggest that $\alpha_{\rm crit}$
does depend on the level of irradiation and is thus 
non-universal. 

The main goal of this work is to explore disk 
characteristics in the gravitoturbulent state, which
can persist for a long time over a 
significant range of radii. For that reason it 
is important to understand how the disk evolves in this regime 
under the action of the non-axisymmetric gravitational torques. 
Currently direct 2D and 3D simulations are too numerically 
expensive to permit such exploration, and one often has to 
resort to azimuthally-averaged, one-dimensional (in radius) 
disk models. To evolve them properly one must provide a 
description of the angular momentum transport by 
the gravitoturbulence. Formulating such a description is
the focus of the present work. 

Balbus \& Papaloizou (1999) argued that due to the long-range 
nature of the gravitational interaction the angular momentum 
transport due to the non-axisymmetric gravitational 
perturbations is inherently non-local. However, 
Gammie (2001) showed that in cold, geometrically thin disks angular 
momentum transport by the gravitational torques can still 
be described as a local process. Lodato \& Rice (2004, 2005)
and Cossins \etal (2009) confirmed this conclusion numerically, 
certainly for the low-mass disks. 

In this work we adopt the latter point of view and will 
explicitly assume the gravitoturbulent transport to be local 
and characterized by the 
effective viscosity parameter $\alpha_{\rm gt}$.  
Different explicit and implicit prescriptions for $\aGI$ have 
been proposed, which can be classified into two general 
categories. 

The first class of $\aGI$ prescriptions relies on the fact that 
in local dynamical and thermal equilibrium the angular momentum 
transport in the disk is intimately related to its thermal state. 
This allows one to directly express $\aGI$ via the disk temperature 
and density. Using constant $\dot M$ disk without external energy 
inputs as an example, 
Rafikov (2009) has demonstrated that this property, when coupled 
with the defining characteristic of the gravitoturbulence --- the
condition $Q\approx Q_0$, allows one to directly relate $\aGI$ to 
$\Sigma$ providing a complete description of the disk evolution. 
This approach has also been implicitly featured in several numerical 
studies (Terquem 2008; Clarke 2009; Zhu \etal 2009a).

A different way of describing the gravitoturbulent disk 
evolution does not explicitly assume $Q\approx Q_0$. Instead, it specifies 
the {\it explicit dependence} of $\aGI$ on $Q$ (Kratter \etal 
2008; Zhu \etal 2009b, 2010a, 2010b; Martin \& Lubow 2011, 2013, 
2014). This approach dates back to the work of Lin \& Pringle 
(1987), who suggested that $\aGI\approx Q^{-2}$ based on a 
phenomenological description of transport in gravitationally 
unstable disks. However, in most cases such prescriptions do
not naturally follow from physical arguments. Instead, they are 
designed to replicate certain qualitative features of $\aGI$ 
behavior. 

The goal of this work is to formulate, analyze and compare 
different approaches to characterizing $\aGI$. First, in 
\S \ref{sect:eqs} we discuss equations used to describe evolution of 
gravitationally unstable disks. Then, in \S \ref{sect:closure} 
we highlight the differences between the two aforementioned 
approaches to closing the system of evolution equations. After
considering the steady, constant $\dot M$ models in \S 
\ref{sect:const_dotM}, we contrast the performance of different
gravitoturbulent viscosity prescriptions in  \S \ref{sect:compare}.
We discuss our results in \S \ref{sect:disc}, where we show, 
in particular, how the gravitoturbulent prescription based on 
$Q\approx Q_0$ condition can be naturally used to obtain 
simple analytical expressions for the
locations of the ice lines and dead zone edges in the 
gravitoturbulent protoplanetary disks.

%%%%%%%%%%%%%%%%%%%%%%%%%%%%%%%%%%%%%%%%%%%%%%%%%%%%%%%%%%%
%%%%%%%%%%%%%%%%%%%%%%%%%%%%%%%%%%%%%%%%%%%%%%%%%%%%%%%%%%%

\section{Basic equations.}  
\label{sect:eqs}

%%%%%%%%%%%%%%%%%%%%%%%%%%%%%%%%%%%%%%%%%%%%%%%%%%%%%%%%%%%

Provided that gravitoturbulent stress can be described
as local $\alpha$-viscosity, it is convenient (Rafikov 2013) to
characterize the disk structure via the angular momentum flux
$F_J$ defined (in a Keplerian disk) as
\ba
F_J\equiv 3\pi\nu\Sigma l=3\pi\alpha c_s^2\Sigma r^2,
\label{eq:F_J}
\ea
where $l=\Omega r^2$ is the specific angular momentum and 
$\nu$ is the kinematic viscosity expressed through the 
dimensionless parameter $\alpha$ as (Shakura \& 
Sunyaev 1973)
\ba
\nu=\alpha \frac{c_s^2}{\Omega}. 
\label{eq:nu}
\ea
As always, we will assume the gas sound speed $c_s$ to be 
determined by the midplane temperature of the disk.
Angular momentum transport is effected both by 
gravitoturbulence and by any other background sources 
of effective viscosity, such as MRI.  

Mass accretion rate through the disk $\dot M$ (defined to be 
positive for inflow) is related to $F_J$ via a simple relation 
(Rafikov 2013)
\ba
\dot M=\frac{\partial F_J}{\partial l}.
\label{eq:dot M}
\ea
Coupled with the continuity equation, this
results in the following evolution 
equation: 
\ba
\frac{\partial \Sigma}{\partial t}=
\frac{1}{2\pi r}\frac{\partial}{\partial r}
\left[\left(\frac{\partial l}{\partial r}\right)^{-1}
\frac{\partial F_J}{\partial r}\right].
\label{eq:ev_gen}
\ea
Expressing $F_J$ via equation (\ref{eq:F_J}) one reproduces 
the classical equation for the viscous disk evolution 
(Lightman \& Eardley 1974; Lin \& Papaloizou 1996)
\ba
\frac{\partial \Sigma}{\partial t}=
\frac{3}{r}\frac{\partial}{\partial r}\left[r^{1/2}
\frac{\partial}{\partial r}\left(r^{1/2}
\nu\Sigma\right)\right].
\label{eq:ev_classical}
\ea
Solving either of these equations requires the 
knowledge of the thermodynamical 
properties of the disk since both $F_J$ and $\nu$ are functions 
of gas temperature $T$, see equations (\ref{eq:F_J}) \& (\ref{eq:nu}).

We now address the thermal state of the disk. We assume 
the disk to be in thermal equilibrium, so that energy sources  
and sinks are in local balance. The former include viscous 
dissipation and external irradiation intercepted by the disk.
The latter is effected by radiative cooling. For local angular 
momentum transport the rate of viscous energy dissipation per 
unit radial distance in the disk is $d\dot E/dr=-F_J d\Omega/dr$. 
Then the rate of energy production by viscous stresses per 
unit area of the disk is
\ba
\dot Q=\frac{1}{2\pi r}\frac{d\dot E}{dr}=\frac{3}{4\pi}
\frac{F_J\Omega}{r^2}.
\label{eq:dotQ}
\ea

Gravitationally unstable disks are often studied in the
limit of a {\it self-luminous} disk (Gammie 2001; Rafikov 2009), 
in which viscous heating of any nature dominates over the energy 
input due to external irradiation. This assumption should be 
reasonable for e.g. Class 0 T Tauri stars, which are 
characterized by intense mass accretion and relatively low 
luminosity of a very young central star, which is also heavily 
obscured. 

However, in general, the disk, especially its outer parts, may 
also be heated appreciably by the external radiation field with temperature 
$T_{\rm irr}(r)$ (Rice \etal 2011; Zhu \etal 2012); $T_{\rm irr}$ 
can be a function of $r$ if irradiation is due to the central 
object. To account for this possibility we describe the local 
thermal balance, which ultimately determines the midplane disk 
temperature $T$, via the following approximate relation:
\ba
\dot Q=\frac{3}{4\pi}\frac{F_J\Omega}{r^2}=\frac{2\sigma}{f(\tau)}
\left(T^4-T_{\rm irr}^4\right).
\label{eq:th_bal}
\ea 
Details of the vertical transport of radiation in the 
disk are specified here via an explicit function $f(\tau)$ of the 
optical depth
\ba
\tau\approx \kappa(T)\Sigma.
\label{eq:tau}
\ea
To zeroth order the disk optical depth is determined predominantly 
by the midplane value of the temperature $T$ ($\kappa$ can also 
be a function of gas density). 

The explicit form of $f(\tau)$ depends on whether the disk is 
optically thick or thin and on the mechanism of the vertical 
energy transport (Rafikov 2007). Here we assume 
radiative energy transfer (Rafikov 2007 has derived 
$f(\tau)$ for convective transport of energy) with dust-dominated
opacity. In this case $f(\tau)$ can be reasonably well approximated 
(up to constant factors of order unity) by
\ba
f(\tau)\approx \tau+\tau^{-1}.
\label{eq:ftau}
\ea
This expression does not discriminate between the Rosseland and 
Planck mean opacities and is approximately valid (up to constant 
factors of order unity) both in the optically thick and thin 
regimes (Goodman 2003). It is easy to see that the condition 
(\ref{eq:th_bal}) supplemented with equation (\ref{eq:ftau})
properly describes the disk thermal balance in
different asymptotic regimes of $\tau$. Indeed, in the optically
thick case ($\tau\gg 1$) one finds 
$\sigma T^4 \approx \sigma T_{\rm irr}^4+(\dot Q/2)\tau$, while in 
the optically thin regime ($\tau\ll 1$) one has 
$\sigma T^4 \approx \sigma T_{\rm irr}^4+(\dot Q/2)\tau^{-1}$.
Both expressions are valid up to constant factors in the 
appropriate limits.

For simplicity in this work we will consider the following 
behavior of $\kappa$ appropriate for dust opacity in certain 
temperature regimes (Bell \& Lin 1994):
\ba
\kappa=\kappa_0 T^\beta,
\label{eq:kappa}
\ea
with $\kappa_0$, $\beta$ being constants. For example, at low 
temperatures, below $170$ K the opacity is dominated by ice 
grains and is characterized by (Bell \& Lin 1994)
\ba
\beta=2,~~~~
\kappa_0=5\times 10^{-4}~{\rm cm^2 ~g^{-1} ~K^{-2}}.
\label{eq:ice-opacity}
\ea
Our results can be 
trivially extended to other, more complicated forms of 
$\kappa$.

Equations (\ref{eq:th_bal})-(\ref{eq:ice-opacity}) provide 
the sought relation between the disk temperature and 
surface density, which is used to determine the viscosity
behavior in equation (\ref{eq:ev_classical}).

Our final note on disk thermodynamics concerns the case
when  cooling is not due to dust emission. In particular,
Latif \& Schleicher (2014a,b) considered the structure of 
gravitationally unstable disks around Pop III stars. Such 
disks are believed to be metal-free (i.e. no dust opacity),
with cooling provided by molecular hydrogen. In this case,
the right-hand side of the energy balance equation 
(\ref{eq:th_bal}) should be modified to $2(c_s/\Omega)\Lambda$,
where $c_s/\Omega$ is the disk scale height and $\Lambda$ is
the volumetric cooling rate. This expression is valid in the
optically thin regime, with certain modifications required
when line cooling switches to the optically thick regime 
(Latif \& Schleicher 2014a; Ripamonti \& Abel 2004).

%%%%%%%%%%%%%%%%%%%%%%%%%%%%%%%%%%%%%%%%%%%%%%%%%%%%%%%%%%%
%%%%%%%%%%%%%%%%%%%%%%%%%%%%%%%%%%%%%%%%%%%%%%%%%%%%%%%%%%%

\section{Closure in the gravitoturbulent state.}  
\label{sect:closure}

%%%%%%%%%%%%%%%%%%%%%%%%%%%%%%%%%%%%%%%%%%%%%%%%%%%%%%%%%%%

Definition (\ref{eq:F_J}) and thermal balance condition 
(\ref{eq:th_bal}) supplemented with equations (\ref{eq:tau}) 
and (\ref{eq:ftau}) allow one to uniquely relate $T$ to $\Sigma$
{\it if the behavior of $\alpha$ is known}. This situation is 
typical for the disk regions dominated by the background viscosity 
(e.g. due to MRI), where one usually knows (or, more often, 
postulates) some behavior of $\alpha$. Knowing the $T(\Sigma)$ 
dependence one can express $F_J$ and $\nu$ as a function 
of $\Sigma$ only, leaving $\Sigma$ as the {\it only dependent 
variable} in the equation (\ref{eq:ev_classical}). This provides 
a {\it closure} of the system of equations for the density, 
angular momentum, and thermal balance, and allows one to 
self-consistently evolve $\Sigma$ in time and space.

In gravitoturbulent state the behavior of $\alpha$ is not known 
a priori and this approach does not apply. To close 
the system of evolution equations one needs to make additional 
assumptions. Two conceptually different approaches to this 
problem are described next.

%%%%%%%%%%%%%%%%%%%%%%%%%%%%%%%%%%%%%%%%%%%%%%%%%%%%%%%%%%%

\subsection{Closure via the $Q=Q_0$ condition.}  
\label{sect:closure1}

Rafikov (2009) pointed out that a necessary closure is naturally 
provided by the the basic property of a gravitoturbulent disk 
following directly from simulations --- that the disk hovers 
on the margin of gravitational stability with 
\ba
Q=Q_0,
\label{eq:Q_Q0}
\ea
where $Q_0\sim 1$. 

Indeed, 
with definition (\ref{eq:Q}) this condition predicts an explicit 
relation between $T$ and $\Sigma$ in the form 
\ba
T_Q(r,\Sigma)=\frac{\mu}{k_{\rm B}}
\left(\frac{\pi G Q_0}{\Omega(r)}\right)^2\Sigma^2.
\label{eq:T_Q}
\ea

Plugging this expression into the thermal balance condition 
(\ref{eq:th_bal}) gives the following expression for the 
gravitoturbulent stress 
\ba
F_{J,{\rm gt}}(r,\Sigma)=\frac{8\pi}{3}\frac{r^2}{\Omega(r)}
\frac{\sigma \left(T_Q^4-T_{\rm irr}^4\right)}{f(\tau)}.
\label{eq:F_J_Sigma}
\ea
The dependence on surface density $\Sigma$ arises here 
because $T_Q$, which 
enters this expression both explicitly and also through 
\ba
\tau(r,\Sigma)=\Sigma\kappa(T_Q(r,\Sigma)), 
\label{eq:tau1}
\ea
is a function of $\Sigma$. 

Combining equations (\ref{eq:F_J})-(\ref{eq:F_J_Sigma}) one also 
comes up with the {\it explicit expression} for the effective 
$\alpha$-parameter due to gravitoturbulence:
\ba
\alpha_{\rm gt}^{\rm R}(r,\Sigma)=\frac{8}{9} 
\frac{\sigma (\pi G Q_0)^6}{f(\tau)}
\left(\frac{\mu}{k_{\rm B}}\right)^4\frac{\Sigma^5}{\Omega^7}
\left(1-\frac{T_{\rm irr}^4}{T_Q^4}\right).
\label{eq:alpha_GI}
\ea
The dependence on $r$ comes only through $\Omega(r)$.
This result was first obtained by Rafikov (2009) in the 
non-irradiated case ($T_{\rm irr}=0$) when studying constant 
$\dot M$ gravitoturbulent disks. However, it clearly 
also holds more generally, for evolving and irradiated 
gravitoturbulent disks, because thermal balance gets 
established faster than the viscous evolution proceeds. 
Note that Rice \etal (2011) found a different (linear) 
scaling with\footnote{Note that the expression in parentheses in 
equation (\ref{eq:alpha_GI}) can be written as 
$\left(1-Q_{\rm irr}^8/Q_0^8\right)$, with 
$Q_{\rm irr}\equiv\Omega c_s\left(T_{\rm irr}\right)/(\pi G\Sigma)$
(Rice \etal 2011).} $T_{\rm irr}$ in their expression for the
gravitoturbulent $\alpha$ because they used a different 
cooling model, namely assuming a constant cooling time.
Expression (\ref{eq:alpha_GI}) for $\alpha_{gt}^{\rm R}$ completes 
the closure and makes evolution equation (\ref{eq:ev_classical}) 
self-consistent.

On the other hand, gravitoturbulent disk evolution can be described 
entirely {\it without introducing} $\alpha$-parameter. Indeed, 
substituting $F_J=F_{J,{\rm gt}}(r,\Sigma)$ given by the expression 
(\ref{eq:F_J_Sigma}) into equation (\ref{eq:ev_gen}) one finds
\ba
\frac{\partial \Sigma}{\partial t}=
\frac{8}{3r}\frac{\partial}{\partial r}\left\{
\frac{1}{\Omega r}\frac{\partial}{\partial r}\left[
\frac{r^2}{\Omega}\frac{\sigma \left(T_Q^4-T_{\rm irr}^4\right)}
{f(\tau(T_Q))}\right]\right\},
\label{eq:ev_gen1}
\ea
where, again, $T_Q=T_Q(r,\Sigma)$, and $T_{\rm irr}(r)$ behavior 
is specified.
This (in general nonlinear) equation is a closed-form, fully 
self-contained evolution equation for $\Sigma$ as a function 
of $t$ and $r$. It represents one of the main results of this 
work.

In the parts of the disk where external irradiation plays the 
dominant role this equation adopts a simple time-independent 
solution $T_Q\to T_{\rm irr}$, so that 
$\Sigma(r)=\Omega(r)c_s\left(T_{\rm irr}(r)\right)/(\pi G Q_0)$
(Rafikov 2009).

%%%%%%%%%%%%%%%%%%%%%%%%%%%%%%%%%%%%%%%%%%%%%%%%%%%%%%%%%%%

\subsubsection{Additional sources of viscosity.}  
\label{sect:add_visc}

Angular momentum transport in the 
disk may be effected not only by gravitoturbulence but also by 
additional stresses. This would be the case, for 
example, if the disk is both sufficiently ionized for the MRI 
to operate and massive enough for being gravitationally unstable. 

To account for the possibility of additional, non-gravitoturbulent 
viscosity parametrized by $\alpha$-parameter $\alpha_m$, one can simply 
write
\ba
F_J &=& \psi F_{J,{\rm gt}}(r,\Sigma)+F_{J,m}(r,\Sigma,T).
\label{eq:F_full}
\ea
Here $F_{J,gt}(r,\Sigma)$ is still given by the expression 
(\ref{eq:F_J_Sigma}) with the switch function $\psi$ introduced
as described below. The non-gravitational viscous angular 
momentum flux is 
\ba
%%%%%%
F_{J,m}(r,\Sigma,T) &=& 3\pi\alpha_m c_s^2(T)\Sigma r^2,
\label{eq:F_m}
\ea
and the behavior of $\alpha_m$ is specified. Thermal balance 
relation (\ref{eq:th_bal}) now gives
\ba
\psi F_{J,{\rm gt}}(r,\Sigma)+F_{J,m}(r,\Sigma,T)=\frac{8\pi}{3}
\frac{r^2}{\Omega}\frac{\sigma \left(T^4-T_{\rm irr}^4\right)}
{f(\tau(\Sigma,T))},
\label{eq:th_bal1}
\ea 
from which we determine $T$ as a function of $r$ and $\Sigma$. 
Then equation (\ref{eq:F_full}) yields $F_J$ as a function of $r$ and 
$\Sigma$, allowing the evolution of $\Sigma$ to be followed 
using equation (\ref{eq:ev_gen}). 

A subtle point in this prescription is that equation 
(\ref{eq:F_full}) {\it always} includes  the gravitoturbulent 
transport in the form (\ref{eq:F_J_Sigma}), with $F_{J,{\rm gt}}$ 
independent of the actual disk temperature $T$ and the 
value of $Q$. As a result, our prescription formally has non-zero 
gravitoturbulent stress $F_{J,{\rm gt}}$ even in the gravitationally 
stable part of the disk, which is not expected. 

For this reason we introduce a switch function $\psi$ in
equation (\ref{eq:F_full}), which takes care of this issue.
There are several different ways in which this can be done.
One method would be to adopt $\psi$ which quickly goes
to zero as soon as $Q\gtrsim Q_0$. However, in the 
absence of a microscopic theory of gravitoturbulence such a
factor would necessarily be introduced in an ad hoc fashion. 

Instead, we have chosen to use
\ba
\psi=\theta(F_{J,{\rm gt}}),
\label{eq:psi}
\ea
where $\theta(z)$ is the Heaviside step function ($\theta(z)=1$ for 
$z\ge 0$; $\theta(z)=0$ for $z< 0$). With this approach
we simply keep $F_{J,{\rm gt}}$ in the form (\ref{eq:F_J_Sigma})
in equation (\ref{eq:F_full}), as long as it is positive. This is 
not a problem, since, as demonstrated in Rafikov (2009), as 
soon as $Q$ exceeds the threshold value 
$Q_0$, the transport is {\it guaranteed to be dominated by the viscous 
stress}, i.e. $F_{J,m}\gg F_{J,{\rm gt}}$. In other words, 
$\alpha_{\rm gt}^{\rm R}\ll \alpha_m$ in gravitationally stable disks,
even if $T_{\rm irr}=0$. And vice versa, $F_{J,m}\ll F_{J,{\rm gt}}$ and 
$\alpha_{\rm gt}^{\rm R}\gg \alpha_m$ when the disk is gravitationally 
unstable and $Q\to Q_0$. This situation is further discussed in \S 
\ref{sect:const_dotM} and is illustrated in
Figure \ref{fig:alpha_trans}, where we display the actual run of the 
viscous and gravitoturbulent transport contributions for a 
particular disk model. 

Under certain circumstances one may find that $T_Q<T_{\rm irr}$ 
in the gravitationally stable part of the disk, so that $F_{J,{\rm gt}}<0$ 
according to equation (\ref{eq:F_J_Sigma}). In this case our
choice (\ref{eq:psi}) of $\psi$ simply guarantees that the 
gravitoturbulent part of the angular momentum flux vanishes 
and $\alpha=\alpha_m$ exactly. 

Evolution equation (\ref{eq:ev_gen}) based on the prescription 
(\ref{eq:F_full}) with $\psi$ given by equation (\ref{eq:psi}) 
allows us to explore 
the transition between the disk regions dominated by 
gravitoturbulence and background viscosity. In the appropriate 
limits its solution reduces to either the gravitoturbulent 
disk solution (for large $r$) following from equation 
(\ref{eq:ev_gen1}), or the conventional viscous disk solution 
(for small $r$, where $F_{J,_{\rm gt}}\ll F_{J,m}$). Alternatively,
one can evolve the disk structure using equation 
(\ref{eq:ev_classical}) with $\nu$ determined by
\ba
\alpha=\psi\alpha_{\rm gt}^{\rm R}(r,\Sigma)+\alpha_m.
\label{eq:alpha_tot}
\ea
It is easy to show that this approach results in almost the same
disk properties, except for the region where 
$\alpha_{\rm gt}^{\rm R}\sim\alpha_m$ and the disk is close to 
marginal gravitational stability.

In summary, the prescription (\ref{eq:F_full})
with $F_{J,{\rm gt}}$ given by (\ref{eq:F_J_Sigma}) should be 
applicable for any value of $Q$ (even though only 
approximately in the parts of the disk just transitioning to 
the {\it marginally} gravitationally unstable regime).

%%%%%%%%%%%%%%%%%%%%%%%%%%%%%%%%%%%%%%%%%%%%%%%%%%%%%%%%%%%

\subsection{Closure via the explicit $\alpha_{\rm gt}(Q)$ dependence.}  
\label{sect:closure2}

An alternative closure scheme that has been broadly used
in the literature {\it postulates} some dependence of 
the gravitoturbulent viscosity $\alpha_{\rm gt}$ on $Q$. 
In this case closure is analogous to a regular viscous 
disk anzatz: equations (\ref{eq:Q}), (\ref{eq:F_J}) with 
$\alpha=\alpha_m+\alpha_{\rm gt}(Q)$, and 
(\ref{eq:th_bal})-(\ref{eq:ftau}) are combined to yield 
a unique 
$T(r,\Sigma)$ relation. It is then plugged into the 
relation (\ref{eq:nu}) for viscosity, resulting in the
expression for $\nu(r,\Sigma)$ and allowing equation 
(\ref{eq:ev_classical}) to be evolved in time. 

Note that the condition (\ref{eq:Q_Q0}) is not used explicitly
in this approach and thus in general there is no guarantee that
this prescription would result in a truly gravitoturbulent disk
structure, with the disk hovering at the edge of instability with
$Q\approx Q_0$. 

In the absence of a microscopic model of gravitoturbulence 
that could motivate a possible dependence of $\alpha_{\rm gt}$ 
on $Q$, several {\it ad hoc} prescriptions for $\alpha_{\rm gt}(Q)$ 
have been suggested. Their
main feature is the rapid increase of $\alpha_{\rm gt}$ as $Q$ 
approaches some critical value $\sim 1$ from above. 

\begin{figure}
\plotone{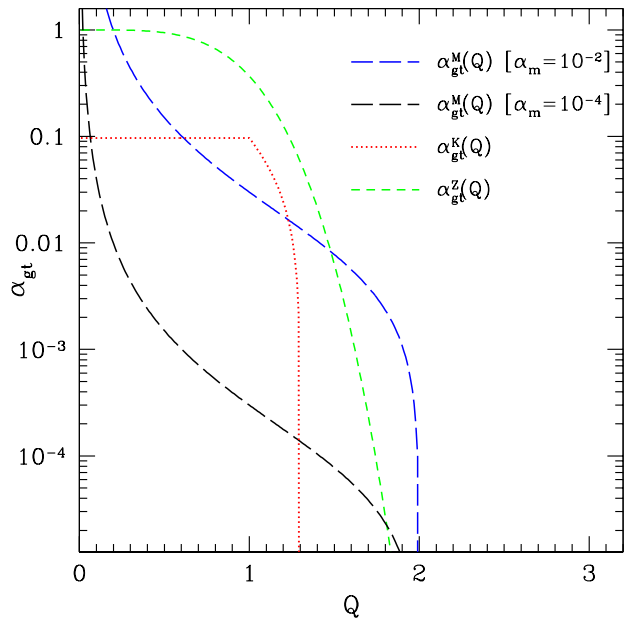}
\caption{
Illustration of different explicit $\alpha_{\rm gt}(Q)$ prescriptions 
given by equations (\ref{eq:Zhu})-(\ref{eq:Kratter}). 
Curves show the runs of $\alpha_{\rm gt}^{\rm Z}(Q)$ of Zhu 
\etal (2010a,b; green, short-dashed), $\alpha_{\rm gt}^{\rm K}(Q)$
of Kratter \etal (2008; red, dotted), equation (\ref{eq:Kratter}), 
and $\alpha_{\rm gt}^{\rm M}(Q)$ of Martin \& Lubow 
(2011; long-dashed), equation (\ref{eq:ML}). The latter is 
displayed for two values of the background viscosity: 
$\alpha_m=10^{-4}$ (black) and $\alpha_m=10^{-2}$ (blue.)
\label{fig:ad_hoc}}
\end{figure}

In particular, Zhu \etal (2010a,b) 
used\footnote{Zhu \etal (2009b) used 
a very similar prescription $\alpha_{\rm gt}(Q)=e^{-Q^2}$.} 
\ba
\alpha_{\rm gt}^{\rm Z}(Q)=e^{-Q^4}.
\label{eq:Zhu}
\ea

Martin \& Lubow (2011,2013,2014) adopted 
\ba
\alpha_{\rm gt}^{\rm M}(Q)=
{\rm max}\left\{\alpha_m\left[\left(\frac{Q_{\rm crit}}{Q}\right)^2-
1\right],0\right\},
\label{eq:ML}
\ea
with $\alpha_m=10^{-2}$ and $Q_{\rm crit}=2$. The $Q^{-2}$ 
dependence is motivated by the work of Lin \& Pringle (1987).
Note that this prescription explicitly relates the value 
of gravitoturbulent viscosity to the background viscosity 
$\alpha_m$. This recipe was also used in Owen \& Jacquet (2014) 
to explore the chemical evolution of a protoplanetary disk 
undergoing accretion outbursts.

Kratter \etal (2008) have used 
\ba
\alpha_{\rm gt}^{\rm K}(Q)={\rm max}\left\{0.14\left[\left(\frac{1.3}
{{\rm max}(Q,1)}\right)^2-1\right],0\right\}.
\label{eq:Kratter}
\ea
One can see that $\alpha_{\rm gt}^{\rm M}$ reduces to 
$\alpha_{\rm gt}^{\rm K}$ if one uses $0.14$ and $1.3$ 
instead of $\alpha_m$ and $Q_{\rm crit}$. Also, $\alpha_{\rm gt}^{\rm K}$
saturates at a constant value $\approx 0.1$ for $Q\le 1$, unlike
$\alpha_{\rm gt}^{\rm M}$, which can be arbitrarily large for
small $Q$. 

These $\alpha_{\rm gt}(Q)$ prescriptions are illustrated 
in Figure \ref{fig:ad_hoc}. We summarize and analyze their 
properties in \S \ref{sect:compare}.

%%%%%%%%%%%%%%%%%%%%%%%%%%%%%%%%%%%%%%%%%%%%%%%%%%%%%%%%%%%
%%%%%%%%%%%%%%%%%%%%%%%%%%%%%%%%%%%%%%%%%%%%%%%%%%%%%%%%%%%

\section{Constant $\dot M$ disks.}  
\label{sect:const_dotM}

%%%%%%%%%%%%%%%%%%%%%%%%%%%%%%%%%%%%%%%%%%%%%%%%%%%%%%%%%%%

One is often interested in knowing the structure of a 
steady-state accretion disk, which necessarily has 
$\dot M$ independent of $r$. This limiting case provides 
a nice point of comparison between the different types 
of viscosity prescriptions and will be used in 
\S \ref{sect:compare}.

If $\dot M$ is independent of $r$ (and, correspondingly, 
$l=\Omega r^2$) and there are no external torques acting 
on the disk, then equation (\ref{eq:dot M}) naturally 
yields $F_J=\dot M l$ (Rafikov 2013). We now illustrate the 
difference in approaches between the various closure 
philosophies when calculating the constant $\dot M$ disk
structure.

%%%%%%%%%%%%%%%%%%%%%%%%%%%%%%%%%%%%%%%%%%%%%%%%%%%%%%%%%%%

\subsection{Constant $\dot M$ disk: $Q=Q_0$ closure}  
\label{sect:closure3}

When we explicitly demand a constant $\dot M$ disk to be 
marginally unstable with the condition (\ref{eq:Q_Q0}),
the equation (\ref{eq:F_full}) implies a relation  
between $\Sigma$, $T$, $r$ in the form
\ba
\psi F_{J,{\rm gt}}(r,\Sigma)+F_{J,m}(r,\Sigma,T)=\dot M l(r). 
\label{eq:cl1}
\ea
The second algebraic relation between these variables is
provided by equations (\ref{eq:th_bal})-(\ref{eq:kappa}), 
thus fully specifying the behavior of $\Sigma(r)$ and 
$T(r)$. Rafikov (2009) and Clarke (2009) used this 
approach to understand 
the properties of gravitoturbulent disks with constant 
$\dot M$.

%%%%%%%%%%%%%%%%%%%%%%%%%%%%%%%%%%%%%%%%%%%%%%%%%%%%%%%%%%%

\subsection{Constant $\dot M$ disk: explicit 
$\alpha_{\rm gt}(Q)$ closure}  
\label{sect:closure4}

When one uses an explicit $\alpha_{\rm gt}(Q)$ prescription, 
the approach to determining disk structure is different 
from that in \S \ref{sect:closure3}. Assuming a 
non-zero background viscosity, equation (\ref{eq:F_J}) 
now yields for the constant $\dot M$ disk
\ba
3\pi c_s^2\Sigma\left[\alpha_{\rm gt}(Q)+\alpha_m\right] 
=\dot M \Omega(r).
\label{eq:cl2}
\ea
Here both $c_s$ and $Q$ explicitly depend on temperature.
Thus, $\alpha_{\rm gt}(Q)$ is a function of $T$ in this 
approach, unlike the case considered in \S \ref{sect:closure1}.

Again, equations (\ref{eq:th_bal})-(\ref{eq:kappa}) provide
a second relation between $\Sigma$, $T$, $r$, allowing one 
to fully determine the constant $\dot M$ disk structure
(see e.g. Zhu \etal 2010a,b; Martin \& Lubow 2011).

%%%%%%%%%%%%%%%%%%%%%%%%%%%%%%%%%%%%%%%%%%%%%%%%%%%%%%%%%%%
%%%%%%%%%%%%%%%%%%%%%%%%%%%%%%%%%%%%%%%%%%%%%%%%%%%%%%%%%%%

\section{Comparison of different $\alpha_{\rm gt}$ prescriptions}  
\label{sect:compare}

%%%%%%%%%%%%%%%%%%%%%%%%%%%%%%%%%%%%%%%%%%%%%%%%%%%%%%%%%%%

We now compare the performance of different prescriptions 
for $\alpha_{\rm gt}$ described 
in \S \ref{sect:closure1}-\ref{sect:closure2}. To that effect 
we construct steady state (constant $\dot M$) models of
gravitoturbulent protoplanetary disks using the results of 
\S \ref{sect:const_dotM}. We then compare the behaviors of 
various disk characteristics obtained with different 
$\alpha_{\rm gt}$ recipes. 

Our calculations adopt the opacity behavior 
(\ref{eq:kappa})-(\ref{eq:ice-opacity}) typical for 
low temperatures $T\lesssim 200$ K
(Bell \& Lin 1994). For simplicity we will assume this behavior 
to extend also to higher temperatures; this should not be a problem 
as out present goal is to explore the {\it differences} 
between the various $\alpha_{\rm gt}$ prescriptions. We take
the mass of the central star to be $M_\star=M_\odot$ and set the 
threshold for gravitational stability at $Q_0=1.5$. In this and 
subsequent calculations we assume that fragmentation
ensues when $\alpha$ exceeds $\alpha_{\rm crit}=0.1$ (rather 
than unity, as e.g. is assumed in equations 
(\ref{eq:sig_f})-(\ref{eq:M_f})). Numerical 
studies suggest that a lower value of $\alpha_{\rm crit}$ 
is more realistic (Rice \etal 2005; Paardekooper \etal 2011). 

Rafikov (2009) showed that gravitoturbulent state results 
in a set of fiducial values of various disk variables: surface 
density $\Sigma_f$, midplane temperature and sound speed $T_f$ 
and $c_{s,f}$, mass accretion rate $\dot M_f$, and
angular frequency $\Omega_f$. They are chosen such 
that at the radius corresponding to the angular frequency $\Omega_f$ a 
steady gravitoturbulent disk with the mass accretion rate equal 
to $\dot M_f$ simultaneously (1) has midplane temperature and 
sound speed equal to $T_f$ and $c_{s,f}$, (2) fragments, and 
(3) has optical depth $\tau=1$. In Appendix \ref{sect:fiducial} we 
provide explicit expressions and numerical estimates for these variables 
for the case of interest to us (opacity $\kappa\propto T^2$). 
These fiducial quantities provide characteristic values of the 
important parameters of gravitoturbulent disk and will be used 
in our subsequent comparisons. 

For a given mass of a central object $M_\star$ characteristic angular 
frequency $\Omega_f$ determines a fiducial distance $r_f$ according
to the formula
\ba
r_f=\left(\frac{G M_\star}{\Omega_f^2}\right)^{1/3}\approx 130~\mbox{AU}
\left(\frac{M_\star}{M_\odot}\right)^{1/3},
\label{eq:r_f}
\ea
where we took the numerical value of $\Omega_f$ from equation 
(\ref{eq:cs_f}) in Appendix \ref{sect:fiducial}. This is the 
radius beyond which an optically thick disk with opacity 
$\kappa\propto T^2$ inevitably fragments (Matzner \& Levin 2005; 
Rafikov 2009; Clarke 2009).

\begin{figure}
\plotone{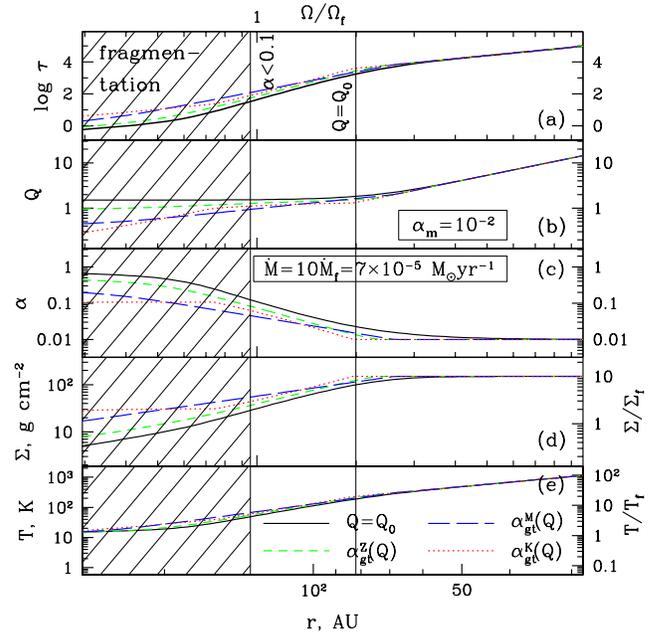}
\caption{Comparison of gravitoturbulent disk models computed 
using $Q=Q_0$ closure (black solid curve) and several 
different $\alpha_{\rm gt}(Q)$ prescriptions: 
$\alpha_{\rm gt}^{\rm Z}(Q)$ of Zhu \etal (2010a,b; green, 
short-dashed), $\alpha_{\rm gt}^{\rm M}(Q)$ of Martin \& Lubow 
(2011; blue, long-dashed), and $\alpha_{\rm gt}^{\rm K}(Q)$
of Kratter \etal (2008; red, dotted). Plotted are:
(a) optical depth, (b) Toomre $Q$, (c) $\alpha$-parameter, (d) 
surface density, (e) midplane temperature. Opacity in the form 
(\ref{eq:kappa})-(\ref{eq:ice-opacity}) is assumed. Background 
viscosity is high, $\alpha_m=10^{-2}$. A high-$\dot M$ case 
is considered ($\dot M=10\dot M_f\approx 7\times 10^{-5}$ 
M$_{\odot}$ yr$^{-1}$) with the disk remaining optically 
thick all the way out to $\approx 150$ AU. Beyond this radius it 
inevitably fragments as $\alpha_{\rm gt}$ exceeds critical value 
$\alpha_{\rm crit}=0.1$. 
\label{fig:high_alpha}}
\end{figure}

In all our models we assume that the disk is immersed in a 
uniform radiation field with the radially-independent temperature
$T_{\rm irr}=T_f\approx 12$ K. We have checked that all our
conclusions remain the same for other values of $T_{\rm irr}$,
in particular for the self-luminous disks with $T_{\rm irr}=0$.

In Figure \ref{fig:high_alpha} we show the radial profiles of the
midplane temperature $T$, surface density $\Sigma$, optical depth
$\tau$, $\alpha$-parameter and Toomre $Q$ for a gravitoturbulent 
disk that has a high mass accretion rate, 
$\dot M=10 \dot M_f\approx 7\times 10^{-5}$ 
M$_{\odot}$ yr$^{-1}$ (leaving aside the question of how realistic 
such $\dot M$ is). We also use a relatively high value of the 
background viscosity, $\alpha_m=10^{-2}$, which is often adopted 
in the literature (Zhu etal 2009a; 
Martin \& Lubow 2011, 2013). Different curves correspond to 
$\alpha_{\rm gt}$ prescriptions given by equations (\ref{eq:alpha_GI}), 
(\ref{eq:Zhu})-(\ref{eq:Kratter}), as indicated in the Figure.

Shaded region outside of $\sim 130$ AU corresponds to a fragmenting
part of the disk where $\alpha>\alpha_{\rm crit}$, and the behavior 
of disk parameters in this region is irrelevant (many curves are 
very discrepant there). As expected (Rafikov 2009), at the 
fragmentation edge our disk model has 
$\Sigma>\Sigma_f$, $T>T_f$, and is optically thick ($\tau\sim 10$).

One can see that all $\alpha_{\rm gt}$ prescriptions reliably 
reproduce the radius at which the transition from the gravitationally 
stable to the gravitoturbulent state occurs (around the vertical 
$Q=Q_0$ line). Interior to this radius background viscosity dominates,
$\alpha\approx \alpha_m$, and different curves overlap. But outside 
this radius, in the region where $\alpha\gtrsim \alpha_m$, 
some differences between $\alpha_{\rm gt}$ prescriptions 
start emerging. In particular, at the fragmentation edge 
(boundary of the shaded region) $\alpha_{\rm gt}^{\rm R}$ 
computed through $Q=Q_0$ closure is different from  
$\alpha_{\rm gt}^{\rm M}$ of Martin \& Lubow (2011) by about a 
factor of $2$. Similar differences at the same 
$r\approx 130$ AU are seen in the values of 
$\Sigma$ and $\tau$. Other explicit $\alpha_{\rm gt}$ prescriptions 
show better agreement (within tens of per cent) for $\Sigma$, 
$\alpha$, and $\tau$ with the 
results of $Q=Q_0$ closure. On the other hand, the behavior of the 
midplane temperature (Figure \ref{fig:high_alpha}e) shows good
agreement between different $\alpha_{\rm gt}$ prescriptions. 
As expected, $T$ never drops below about 10 K as it is limited by 
$T_{\rm irr}$ far from the star; however, for this disk model this 
is true only beyond the fragmentation edge so that in practice 
irradiation with the adopted $T_{\rm irr}$ is irrelevant here.

\begin{figure}
\plotone{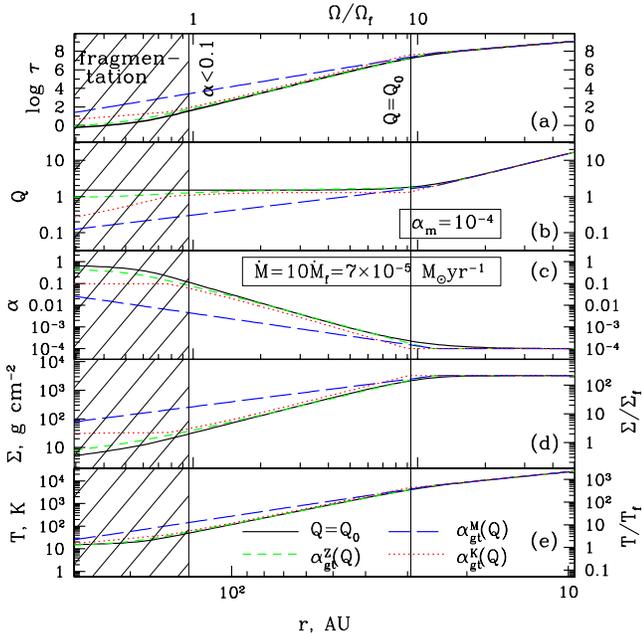}
\caption{
Same as Figure \ref{fig:high_alpha} but now for the low background
viscosity $\alpha_m=10^{-4}$. One can see significant differences 
between the results obtained using $\alpha_{\rm gt}^{\rm M}(Q)$ recipe 
and all other $\alpha_{\rm gt}$ prescriptions, which tend to generally 
agree with each other.
\label{fig:high_Mdot}}
\end{figure}

To investigate these differences further we recomputed the disk
structure for the same high mass accretion rate $7\times 10^{-5}$
M$_\odot$ yr$^{-1}$ but decreasing the background viscosity
by two orders of magnitude, to $\alpha_m=10^{-4}$. This lower 
value of $\alpha_m$ may be 
more realistic for the cold protoplanetary disks, where the MRI 
is weakened by the non-ideal effects such as ambipolar diffusion 
(Bai \& Stone 2011), or may not be operating at all, e.g. in a 
dead zone (Gammie 1996;	Fleming \& Stone 2003).

Figure \ref{fig:high_Mdot} shows the results of this calculation.
Its inspection reveals good quantitative agreement between the 
models computed via $Q=Q_0$ closure (black solid), which rely on 
$\alpha_{\rm gt}^{\rm R}$ given by the equation (\ref{eq:alpha_GI}), 
and those using explicit viscosity in the form proposed by 
Kratter \etal (2008),
$\alpha_{\rm gt}^{\rm K}(Q)$, and Zhu \etal (2010a,b),
$\alpha_{\rm gt}^{\rm Z}(Q)$. They typically agree
with several tens of per cent for all variables in the 
whole gravitoturbulent region of the disk, between the 
fragmentation edge and the $Q=Q_0$ line.  

It is also obvious that the use of $\alpha_{\rm gt}^{\rm M}(Q)$ 
suggested by Martin \& Lubow (2011) and $\alpha_m$ this low leads 
to rather poor quantitative agreement with all other 
prescriptions: $\Sigma$, $\alpha$ and $\tau$
show a discrepancy of more than an order of magnitude in the outer 
disk, at the edge of the fragmentation zone. The value of $Q$ at
this radius plunges to $\sim 0.2$ for $\alpha_{\rm gt}^{\rm M}(Q)$, 
in contrast to $Q\approx Q_0$ maintained by all other prescriptions
in agreement with simulations. Lower value of $\alpha_m$ used in 
this calculation pushes gravitoturbulent zone down to 
$\sim 30$ AU, accentuating the deviations which 
were already visible at higher $\alpha_m$, with less extended 
gravitoturbulent region (see Figure \ref{fig:high_alpha}).

\begin{figure}
\plotone{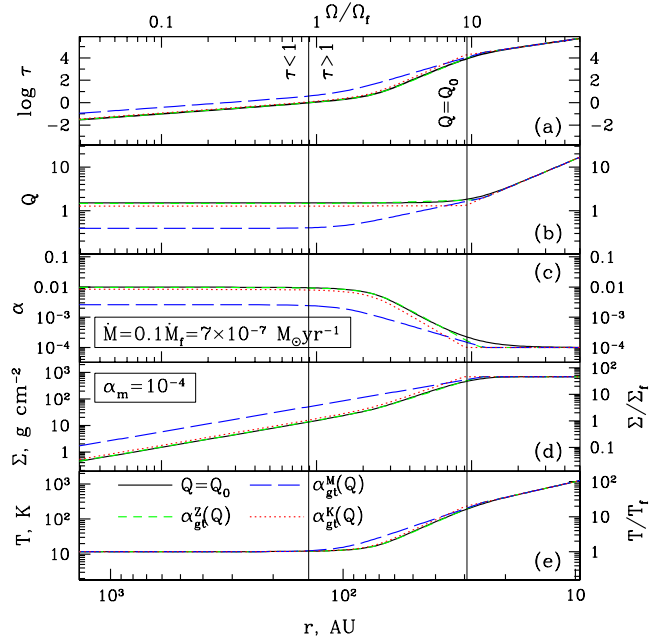}
\caption{
Same as Figure \ref{fig:high_Mdot} (same low $\alpha_m=10^{-4}$) 
but now for the low-$\dot M$ case, 
$\dot M=0.1\dot M_f\approx 7\times 10^{-7}$ M$_{\odot}$ yr$^{-1}$. 
Disk becomes optically thin around $10^2$ AU and remains stable 
against fragmentation out to $\gtrsim 10^3$ AU.
\label{fig:low_Mdot}}
\end{figure}

This general situation holds for other disk models. For example, 
in Figure 
\ref{fig:low_Mdot} we compare different $\alpha_{\rm gt}$ recipes
in a low-$\dot M$ disk with $\dot M=0.1\dot M_f=7\times 10^{-7}$ 
M$_\odot$ yr$^{-1}$, while keeping $\alpha_m=10^{-4}$. 
This disk is optically thin in its outer parts, with $\tau=1$ transition
happening {\it interior} to $r_f$. In this case, as shown in Rafikov 
(2009), fragmentation can be pushed out to very large radii and 
the disk can extend beyond $10^3$ AU in the optically thin, 
gravitoturbulent regime. Moreover, external irradiation is very important 
in this case: $T_{\rm irr}\approx 12$ K sets the temperature floor 
outside 100 AU, in the part of the disk which is stable against 
fragmentation, see Figure \ref{fig:low_Mdot}e.

Again, with such low $\alpha_m$ disk models using 
$\alpha_{\rm gt}^{\rm M}(Q)$ suggested by 
Martin \& Lubow (2011) are considerably different from all other 
models, which tend to agree with each other. For $\Sigma$ the 
discrepancy can be as high 
as an order of magnitude. From this we can conclude that 
$\alpha_{\rm gt}^{\rm M}(Q)$ prescription underperforms at low values 
$\alpha_m$. We discuss the reasons for this next.

%%%%%%%%%%%%%%%%%%%%%%%%%%%%%%%%%%%%%%%%%%%%%%%%%%%%%%%%%%%
%%%%%%%%%%%%%%%%%%%%%%%%%%%%%%%%%%%%%%%%%%%%%%%%%%%%%%%%%%%

\section{Discussion.}  
\label{sect:disc}

%%%%%%%%%%%%%%%%%%%%%%%%%%%%%%%%%%%%%%%%%%%%%%%%%%%%%%%%%%%

Our interpretation of the behavior found in models using 
$\alpha_{\rm gt}^{\rm M}(Q)$ is that it is due to the rather 
gradual dependence of the former on $Q$. Indeed, Figure 
\ref{fig:ad_hoc} shows that when 
$\alpha_{\rm gt}$ is between $\alpha_m$ and the critical value 
for fragmentation ($\alpha_{\rm crit}=0.1$ in our case) 
$\alpha_{\rm gt}^{\rm M}(Q)$
grows with decreasing $Q$ much slower than either 
$\alpha_{\rm gt}^{\rm K}(Q)$ or $\alpha_{\rm gt}^{\rm Z}(Q)$. 
This is especially obvious in the $\alpha_m=10^{-4}$ case, when 
$Q$ has to go down to $Q\approx 0.1$ for $\alpha$ to reach 
$\alpha_{\rm crit}$. This explains a significant drop in $Q$ 
towards the fragmentation edge in Figures \ref{fig:high_Mdot}b
and \ref{fig:low_Mdot}b. Thus, we conclude that 
$\alpha_{\rm gt}^{\rm M}(Q)$ does not properly capture the main 
feature of the gravitoturbulent disk --- almost constant and 
close to unity value of the Toomre $Q$. The explicit dependence 
of $\alpha_{\rm gt}^{\rm M}(Q)$ on $\alpha_m$ may be another 
feature that causes it to underpeform compared to 
$\alpha_{\rm gt}^{\rm K}(Q)$ and $\alpha_{\rm gt}^{\rm Z}(Q)$.

At the same time, we must emphasize that we are not aware 
of the existing gravitoturbulent disk calculations using 
low $\alpha_m=10^{-4}$ --- most of them assume 
$\alpha_m=10^{-2}$. Thus, Figures 
\ref{fig:high_Mdot} and \ref{fig:low_Mdot} are not providing 
comparison with any published results, but merely {\it urging 
caution} in choosing $\alpha_{\rm gt}$ recipe when working 
with low $\alpha_m$. And at higher $\alpha_m=10^{-2}$ calculations 
using $\alpha_{\rm gt}^{\rm M}(Q)$ can still be acceptable 
(see Figure \ref{fig:high_alpha}), depending on the required 
accuracy.

Explicit prescriptions suggested in Kratter \etal (2008) and 
Zhu \etal (2010a,b) ($\alpha_{\rm gt}^{\rm K}(Q)$ and 
$\alpha_{\rm gt}^{\rm Z}(Q)$) do rather well at reproducing 
the $Q=Q_0$ calculation using $\alpha_{\rm gt}^{\rm R}$. Based
on this we expect that effectively any explicit 
$\alpha_{\rm gt}(Q)$ scheme that exhibits {\it very sharp rise} 
in the vicinity of $Q_0$ as $Q$ decreases, should be suitable 
for practical calculations of the gravitoturbulent disk 
properties. In fact, the sharper is the better and something 
as simple as $\alpha_{\rm gt}(Q)=\exp[(Q_0-Q)/\delta Q]$ with 
$\delta Q\lesssim 10^{-2}$ will maintain $Q\approx Q_0$ 
in the gravitoturbulent state quite well (conditional upon 
$\alpha<\alpha_{\rm crit}$).

Nevertheless, we strongly believe that there are significant 
benefits to using the $Q=Q_0$ approach outlined in \S 
\ref{sect:closure1}. First, the only assumption that it employs 
is that the disk is able to maintain itself in a state of marginal
gravitational stability $Q=Q_0$. This stipulation is in agreement 
with the results of direct numerical simulations (Cossins \etal 
2009, 2010). Explicit $\alpha_{\rm gt}(Q)$ prescriptions 
also make this assumption, even though implicitly. But in 
addition they have to postulate some actual 
functional form of the $\alpha_{\rm gt}(Q)$ dependence, which 
does not have physical justification and is always introduced 
in an ad hoc fashion. This makes the explicit 
$\alpha_{\rm gt}(Q)$ approach rather arbitrary, which may result
in certain problems, as we have demonstrated in \S 
\ref{sect:compare}. 

Second, deep in the gravitoturbulent regime, when one can neglect 
the background viscosity $\alpha_m$ compared to the gravitoturbulent
contribution $\alpha_{\rm gt}^{\rm R}$, the explicit $Q=Q_0$ approach 
results in a self-contained equation (\ref{eq:ev_gen1}) for 
the surface density evolution. This is allowed by the fact that the
expression (\ref{eq:alpha_GI}) already {\it implicitly accounts for
the thermal balance of the disk}, even in presence of external
irradiation. Disk temperature is uniquely 
defined in this case by the relation (\ref{eq:T_Q}) {\it after 
the disk surface density has been solved for}. This is not the 
case for the explicit $\alpha_{\rm gt}(Q)$ recipes, which always 
need to {\it explicitly relate $T$ to $\Sigma$ via the equation 
of thermal balance} (even deep in the gravitoturbulent state, when
$\alpha_{\rm gt}\gg \alpha_m$), because both enter the definition 
(\ref{eq:Q}) of the Toomre $Q$. Thus, our $Q=Q_0$ 
prescription allows a more efficient numerical exploration of 
the gravitoturbulent disk structure.

Third, when the background viscosity cannot be neglected compared
to $\alpha_{\rm gt}$, the explicit $Q=Q_0$ prescription 
(\ref{eq:alpha_GI}) does not make the task of computing the 
disk properties more complicated than the explicit 
$\alpha_{\rm gt}(Q)$ prescriptions, see \S \ref{sect:add_visc}.
Dealing with this case using our prescription (\ref{eq:alpha_tot}), 
which results in a non-zero $\alpha_{\rm gt}$ even 
in the gravitationally stable part of the disk, is not 
a problem. 

\begin{figure}
\plotone{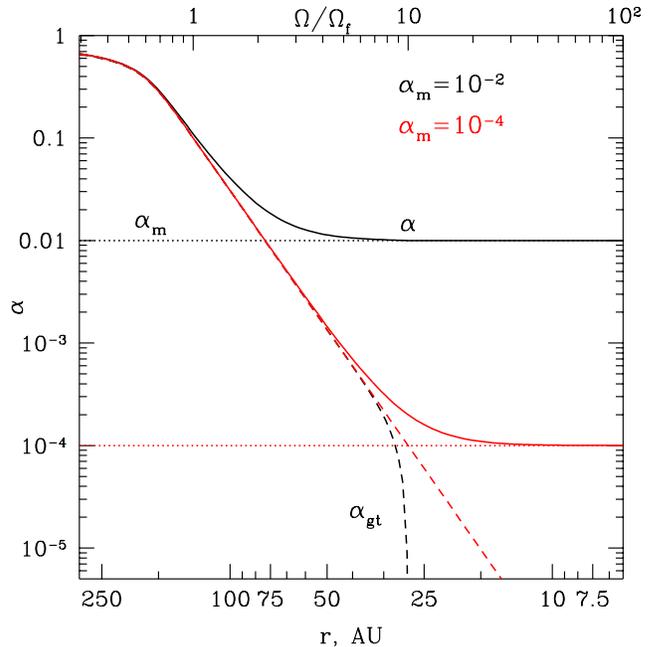}
\caption{
Behavior of $\alpha$ at the transition between the 
gravitationally stable and unstable disk regions. Curves
of $\alpha$ (solid), $\alpha_{\rm gt}^{\rm R}$ (dashed), $\alpha_m$ 
(dotted) are shown as functions of $r$ (lower axis) and 
$\Omega$ (upper axis) for $\alpha_m=10^{-2}$ (black) and 
$\alpha_m=10^{-4}$ (red). In both cases 
$\dot M=10\dot M_f\approx 7\times 10^{-5}$ M$_{\odot}$ yr$^{-1}$,
and these results correspond to Figures \ref{fig:high_alpha}
and \ref{fig:high_Mdot}.  
One can see that in the gravitationally 
stable part of the disk (at smaller $r$ and larger $\Omega$) 
$\alpha_{\rm gt}^{\rm R}$ rapidly becomes subdominant compared to 
the background viscosity $\alpha_m$ and $\alpha\to\alpha_m$, 
thus justifying our simple prescription (\ref{eq:F_full}).
\label{fig:alpha_trans}}
\end{figure}

Indeed, in Figure \ref{fig:alpha_trans} we demonstrate 
that $\alpha_{\rm gt}$ computed in our disk models via equation 
(\ref{eq:alpha_GI}) rapidly becomes subdominant as
the disk transitions into the regime dominated by the background
viscosity $\alpha_m$. Thus, even though in general $\alpha_{\rm gt}^{\rm R}$
is not precisely zero\footnote{Graviturbulent contribution does
vanish inside of 25 AU in Figure \ref{fig:alpha_trans} for 
$\alpha_m=10^{-2}$ because of our use of $\psi$ in the form 
(\ref{eq:psi}) and the fact that $T_Q$ drops below 
$T_{\rm irr}$ inside this radius.} in this part of the disk, 
it still does not affect the disk properties. The only 
place where our approach may be 
quantitatively less accurate is around the transition 
$\alpha_{\rm gt}\sim \alpha_m$. However, this is the location 
where all $\alpha_{\rm gt}$ prescriptions have 
some degree of arbitrariness. 

Fourth, $Q=Q_0$ prescription
allows a much more transparent analytical derivation of certain
gravitoturbulent disk characteristics. We demonstrate this next 
when we compare the derivations of the snow line location 
(\S \ref{sect:ice}) and of the dead zone edge 
(\S \ref{sect:deadzone}) in gravitoturbulent disks using the 
two classes of $\alpha_{\rm gt}$ prescriptions.

%%%%%%%%%%%%%%%%%%%%%%%%%%%%%%%%%%%%%%%%%%%%%%%%%%%%%%%%%%%

\subsection{Snow line location}  
\label{sect:ice}

%%%%%%%%%%%%%%%%%%%%%%%%%%%%%%%%%%%%%%%%%%%%%%%%%%%%%%%%%%%

Snow (or ice) lines for different volatile materials (H$_2$O, 
CO, etc.) arise in protoplanetary disks at the locations where 
these materials undergo a transition from solid to gaseous 
phase. It is conventional to identify snow lines with the disk 
radii where the midplane temperature $T$ is equal to some 
characteristic value $T_{\rm snow}$ for sublimation of a 
particular material at the typical midplane pressure. 
Following common wisdom we will adopt this as a working 
definition of a snow line.

Martin \& Livio (2013) derived an analytical expression
for the snow line radius $R_{\rm snow}$ in a gravitoturbulent,
optically thick, self-luminous protoplanetary disk with 
constant $\dot M$. Here we compare their calculation with 
the derivation of $R_{\rm snow}$ using our favored 
$\alpha_{\rm gt}$ recipe (\ref{eq:alpha_GI}) based on 
the condition $Q=Q_0$. To facilitate comparison we 
employ a common set of assumptions: (1) neglect background 
viscosity $\alpha_m$ and external irradiation
(i.e. $T_{\rm irr}=0$), (2) describe optically thick radiation 
transfer by\footnote{This is what equations (4) \& (5) of 
Martin \& Livio (2013) effectively reduce to.} 
\ba
f(\tau)=\frac{3}{8}\tau
\ea
in place of our equation (\ref{eq:ftau}), with 
$\tau=(1/2)\kappa\Sigma$ (instead of our definition 
(\ref{eq:tau1})), and (3) take opacity at the 
icy grain sublimation radius to be given by 
equation (\ref{eq:kappa}) with $\beta=-0.01$ and 
$\kappa_0=3$ (in appropriate CGS units).

%%%%%%%%%%%%%%%%%%%%%%%%%%%%%%%%%%%%%%%%%%%%%%%%%%%%%%%%%%%

\subsubsection{$R_{\rm snow}$ via the $Q=Q_0$ prescription}  
\label{sect:ice1}

In our approach we first express $\Sigma$ via $R_{\rm snow}$ and 
$T_{\rm snow}$ by setting $T_Q=T_{\rm snow}$ in equation 
(\ref{eq:T_Q}). Then we plug this $\Sigma(R_{\rm snow})$ into
the expression (\ref{eq:alpha_GI}) for $\alpha_{\rm gt}^{\rm R}$.
Finally, inserting all this into equation (\ref{eq:cl1}), written 
in the form 
$3\pi c_s^2\Sigma\alpha_{\rm gt}^{\rm R}=\dot M \Omega$ with all 
variables expressed through $R_{\rm snow}$ and 
$T_{\rm snow}$, and resolving it for $R_{\rm snow}$
we get
\ba
R_{\rm snow} &=& \left[\frac{9}{128\pi^2}
\frac{\kappa_0M^{3/2}}{\sigma Q_0}\left(\frac{Gk_{\rm B}}{\mu}\right)
\dot M T_{\rm snow}^{\beta-7/2}\right]^{2/9}
\nonumber\\
%%%%%%%%%
& \approx & 5.78~\mbox{AU}~Q_0^{-2/9}
\left(\frac{\dot M}{\rm 10^{-8}M_\odot yr^{-1}}\right)^{2/9}
\nonumber\\
& \times & \left(\frac{M}{\rm M_\odot}\right)^{1/3}
\left(\frac{T_{\rm snow}}{\rm 170~K}\right)^{-0.78}.
\label{eq:Rsnow}
\ea
Note that this derivation does not use thermodynamic balance 
condition (\ref{eq:th_bal}) as it has already been used in 
deriving the expression (\ref{eq:alpha_GI}) for 
$\alpha_{\rm gt}^{\rm R}$.

%%%%%%%%%%%%%%%%%%%%%%%%%%%%%%%%%%%%%%%%%%%%%%%%%%%%%%%%%%%

\subsubsection{$R_{\rm snow}$ via the explicit $\alpha_{\rm gt}(Q)$}  
\label{sect:ice2}

Now we outline the steps used in Martin \& Lubow (2013) 
to derive $R_{\rm snow}$. They adopted an explicit 
$\alpha_{\rm gt}(Q)=\alpha_0\exp(-Q^4)$
with\footnote{Martin \& Livio (2013) used $\alpha_m$ rather 
than $\alpha_0$ to denote $\alpha_{\rm gt}(0)$ in their 
equation (9) for $\alpha_{\rm gt}(Q)$, 
but this notation would cause confusion with the background viscosity 
in our paper.} $\alpha_0=10^{-2}$. This expression is similar to 
that in Zhu \etal (2009b; see our equation (\ref{eq:Zhu})) 
in spirit, but not in detail, as the maximum value of 
$\alpha_{\rm gt}(Q)$ is modulated by a small factor $\alpha_0$
in the Martin \& Livio (2013) case. Note that this anzatz 
would never allow a gravitationally unstable disk to fragment
unless the critical value of $\alpha_{\rm gt}$ for fragmentation 
$\alpha_{\rm crit}$ were very low, below $\alpha_0$.

With this prescription the constant $\dot M$ assumption, embodied 
in the equation (\ref{eq:cl2}) allowed Martin \& Livio (2013) 
to relate $Q$ at the snow line to the disk midplane temperature 
(assumed equal to $T_{\rm snow}$) and $R_{\rm snow}$. Because of 
the adopted form of $\alpha_{\rm gt}(Q)$ this relation depended 
on $\alpha_0$ and involved a non-elementary function (Lambert 
function), complicating the analysis. The latter problem was 
overcome by using an asymptotic relation for the Lambert 
function, valid for relatively low values of $\dot M\lesssim 
10^{-7}$ M$_\odot$ yr$^{-1}$. 

Additional (algebraic) relation between $T_{\rm snow}$, 
$R_{\rm snow}$, and $\Sigma$ at the snow line comes from the 
thermodynamical relation (\ref{eq:th_bal}) with 
$F_J=\dot M \Omega r^2$. Eliminating $\Sigma(R_{\rm snow})$ with 
the aid of these two relations Martin \& Lubow (2013) obtained 
a scaling for $R_{\rm snow}$ in terms of powers of 
$T_{\rm snow}$, $\dot M$, and the central star mass $M$, given 
by the equation (19) of their paper.

%%%%%%%%%%%%%%%%%%%%%%%%%%%%%%%%%%%%%%%%%%%%%%%%%%%%%%%%%%%

\subsubsection{Comparison of approaches}  
\label{sect:ice3}

Comparing the result for $R_{\rm snow}$ in Martin \& Lubow (2013) 
with equation (\ref{eq:Rsnow}) shows 
that the two are essentially identical, with all the scalings 
being the same and the numerical estimates different at $1\%$ 
level for $Q_0=1$. However, it is clear that our derivation of 
the expression (\ref{eq:Rsnow}) is much more straightforward 
and flexible. 

First, it involves only elementary functions. Second, it does not  
involve ad hoc factors, such as $\alpha_0$ used by Martin \& 
Lubow (2013) in their definition of $\alpha_{\rm gt}(Q)$, which 
are confusing and do not appear 
in the final result anyway. Third, our result (\ref{eq:Rsnow})
is clearly valid regardless of the value of $\dot M$, while 
the asymptotic representation of the Lambert function used by 
Martin \& Lubow (2013) works only for relatively low 
$\dot M\lesssim 10^{-7}$ M$_\odot$ yr$^{-1}$. 

Analogous problems would arise if one were to use 
other explicit $\alpha_{\rm gt}(Q)$ prescriptions
for analytical calculations. This provides
strong motivation for adopting the $Q=Q_0$ approach advocated 
in this work rather than the explicit $\alpha_{\rm gt}(Q)$
recipes, when studying gravitoturbulent disks.

%%%%%%%%%%%%%%%%%%%%%%%%%%%%%%%%%%%%%%%%%%%%%%%%%%%%%%%%%%%
\subsection{Outer edge of the dead zone}  
\label{sect:deadzone}

%%%%%%%%%%%%%%%%%%%%%%%%%%%%%%%%%%%%%%%%%%%%%%%%%%%%%%%%%%%

In a very similar vein we can estimate the location of the 
outer edge of a dead zone (Gammie 1996) in a gravitoturbulent 
protoplanetary disk. This edge should be located at the 
radius $R_{\rm d}$, where the disk surface density is twice 
the surface density of the active layer 
$\Sigma_{\rm a}\approx 100$ g cm$^{-2}$, 
down to which external ionizing radiation can penetrate into 
the disk and keep it fully ionized. 

For this estimate the equation (\ref{eq:T_Q}) of the $Q=Q_0$ 
approach allows one to express the midplane temperature 
through $R_{\rm d}$ and $\Sigma_{\rm a}$, while 
$\alpha_{\rm gt}^{\rm R}$ is already a function of the two,
see equation (\ref{eq:alpha_GI}). Then, repeating the 
steps in \S \ref{sect:ice1} and eliminating $T(R_{\rm d})$ one
obtains
\ba
R_{\rm d} &=& \left(GM\right)^{1/3}
\left[\frac{9}{128\pi}\frac{\kappa_0\dot M}{\sigma \Sigma_a^{7-2\beta}}
\frac{\left(k_{\rm B}/\mu\right)^{4-\beta}}{\left(\pi G Q_0\right)^{2(4-\beta)}}
\right]^{1/(15-3\beta)}
\nonumber\\
%%%%%%%%%
& \approx & 21~\mbox{AU}~Q_0^{-4/9}
\left(\frac{\dot M}{\rm 10^{-8}M_\odot yr^{-1}}\right)^{1/9}
\nonumber\\
& \times & \left(\frac{M}{\rm M_\odot}\right)^{1/3}
\left(\frac{\Sigma_{\rm a}}{\rm 10^2 g~cm^{-2}}\right)^{-1/3}.
\label{eq:Sig_d}
\ea
In making the estimate we again adopted the opacity in 
the form (\ref{eq:ice-opacity}), which is perfectly 
reasonable since $R_{\rm d}>R_{\rm snow}$ and ice grains 
dominate opacity.

It is easy to see that deriving $R_{\rm d}$
via some explicit $\alpha_{\rm gt}(Q)$ prescription would again
require going through the rather contrived procedure described
in \S \ref{sect:ice2}, involving non-elementary functions, 
their asymptotic expansions, and so on. Thus, similar to
\S \ref{sect:ice3}, we can again argue that the $Q=Q_0$ 
approach provides the shortest and clearest path to 
finding $R_{\rm d}$.

%%%%%%%%%%%%%%%%%%%%%%%%%%%%%%%%%%%%%%%%%%%%%%%%%%%%%%%%%%%
%%%%%%%%%%%%%%%%%%%%%%%%%%%%%%%%%%%%%%%%%%%%%%%%%%%%%%%%%%%

\section{Conclusions.}
\label{sect:concl}

%%%%%%%%%%%%%%%%%%%%%%%%%%%%%%%%%%%%%%%%%%%%%%%%%%%%%%%%%%%

One of the main outcomes of this work is the development 
of a self-consistent analytical formalism for the evolution 
of a gravitoturbulent 
accretion disk, including the 
possibility of external irradiation. It 
explicitly enforces Toomre $Q$ in the disk to stay 
close to the value $Q_0$ needed for marginal gravitational 
stability. Implicit numerical implementations of this 
approach do exist (e.g. Terquem 2008; Clarke 2009; Zhu 
\etal 2009a) but a complete analytical formalism was lacking 
until now. Our work fills this gap with the prescriptions 
described in \S \ref{sect:closure1}, in particular
the explicit expression (\ref{eq:alpha_GI}) for the 
gravitoturbulent viscosity $\alpha_{\rm gt}^{\rm R}$, which 
depends only on $r$ and $\Sigma$ and fully accounts for 
the thermodynamic equilibrium of the disk. Our formalism 
includes the possibility of the non-zero background viscosity 
$\alpha_m$ (\S \ref{sect:add_visc}) provided by sources 
other than the gravitoturbulence (e.g. MRI), and external
irradiation of the disk. This work thus generalizes the 
models of steady, constant $\dot M$ gravitoturbulent 
disks explored in Rafikov (2009).

In fully gravitoturbulent state, when 
$\alpha_{\rm gt}^{\rm R}\gg \alpha_m$ we derive an 
{\it explicit equation} (\ref{eq:ev_gen1}) for the 
surface density evolution, which is fully self-contained ---
it involves {\it only} $r$ and $\Sigma$ --- and is 
non-linear in general. This equation provides a powerful 
tool for exploring gravitoturbulent disk evolution 
and is thus an important result of our work. 

We then contrasted our $Q=Q_0$ approach with the calculations 
specifying an {\it explicit dependence} of the 
gravitoturbulent viscosity 
$\alpha_{\rm gt}$ on the Toomre $Q$, which is often 
done in the literature. We clearly demonstrated that our 
$Q=Q_0$ approach provides a much faster and more natural 
route to deriving analytical expressions for various 
important characteristics of the gravitoturbulent disks, 
such as the locations of 
the snow lines (\S \ref{sect:ice}) and the edge of the 
dead zone (\ref{sect:deadzone}). We also compared disk 
models computed using different $\alpha_{\rm gt}$ 
recipes (\S \ref{sect:compare}) and found that some explicit 
$\alpha_{\rm gt}(Q)$ prescriptions are able to reproduce 
the results obtained using $Q=Q_0$ approach reasonably well. 
This is the case 
only for those prescriptions that cause $\alpha_{\rm gt}(Q)$ 
to increase very steeply as $Q$ decreases in the vicinity 
of $Q_0$ (e.g. Kratter \etal 2008; Zhu \etal 2010a,b). 
Certain disagreement may arise if this requirement 
is violated (Martin \& Lubow 2010), especially when the 
background (non-gravitoturbulent) viscosity in the disk 
is low (see our discussion in \S \ref{sect:disc}). Thus, 
although it is often stated in the literature that the
precise form of the $\alpha_{\rm gt}(Q)$ dependence used 
for building gravitoturbulent disk models is unimportant, 
we find this to be only partly true.

Finally, we emphasize that our formalism has the minimal 
number of imposed assumptions --- namely, that $Q=Q_0$ is
approximately maintained in the gravitoturbulent state,
in agreement with simulations. It thus provides a
natural and robust pathway to building efficient 
time-dependent models of gravitoturbulent disks around 
objects such as young stars, quasars, and Pop III stars.

%%%%%%%%%%%%%%%%%%%%%%%%%%%%%%%%%%%%%%%%%%%%%%%%%%%%%%%%%%%

\acknowledgements 

I am grateful to Zhaohuan Zhu for useful comments. 
The financial support for this work is provided by the NSF 
grant AST-1409524.

\appendix

%%%%%%%%%%%%%%%%%%%%%%%%%%%%%%%%%%%%%%%%%%%%%%%%%%%%%%%%%%%
%%%%%%%%%%%%%%%%%%%%%%%%%%%%%%%%%%%%%%%%%%%%%%%%%%%%%%%%%%%

\section{Characteristic disk parameters}  
\label{sect:fiducial}

%%%%%%%%%%%%%%%%%%%%%%%%%%%%%%%%%%%%%%%%%%%%%%%%%%%%%%%%%%%

For our adopted opacity behavior 
(\ref{eq:kappa})-(\ref{eq:ice-opacity}), assuming that fragmentation 
happens at $\alpha_{\rm crit}=1$, the fiducial values of the surface 
density $\Sigma_f$, midplane temperature and sound speed $T_f$ 
and $c_{s,f}$, mass accretion rate $\dot M_f$, and angular 
frequency $\Omega_f$  are
\ba
\Sigma_f & = & \left[\left(\frac{8}{9}\frac{\sigma}{\pi G Q_0}\right)^4
\left(\frac{\mu}{k_{\rm B}}\right)^2\kappa_0^{-7}\right]^{1/15}
\approx 14~\mbox{g}~\mbox{cm}^{-2},
\label{eq:sig_f}
~~~~~~~~
%\\
%& \approx & 14~\mbox{g}~\mbox{cm}^{-2},
%\nonumber\\
%%%%%%%%%%%%%%%%
T_f 
%& 
= 
%&  
\left[\left(\frac{9}{8}
\frac{\pi G Q_0}{\sigma\kappa_0^{2}}\right)^2
\frac{k_{\rm B}}{\mu}\right]^{1/15}\approx 12~\mbox{K},
\label{eq:T_f}\\
%%%%%%%%%%%%%%%%
c_{s,f} & = &  \left[\frac{9}{8}
\frac{\pi G Q_0}{\sigma\kappa_0^{2}}
\left(\frac{\mu}{k_{\rm B}}\right)^{-8}\right]^{1/15}
\approx 0.22~\mbox{km}~\mbox{s}^{-1},
\label{eq:cs_f}
~~~~~~~~
%\\
%%%%%%%%%%%%%%%%
\Omega_f 
%& 
= 
%&  
\left[\frac{8}{9}
\frac{\sigma}{\kappa_0}
\left(\pi G Q_0\frac{\mu}{k_{\rm B}}\right)^2\right]^{1/3}
\approx   1.4\times 10^{-10}~\mbox{s}^{-1},
%\label{eq:omega_f}
\\
%& \approx &  1.4\times 10^{-10}~\mbox{s}^{-1},
%\nonumber\\
%%%%%%%%%%%%%%%%
\dot M_f & = &  3\pi
\left[\frac{8}{9}\sigma\kappa_0^2
(\pi G Q_0)^4
\left(\frac{\mu}{k_{\rm B}}\right)^8\right]^{-1/5}
\approx   7.2\times 10^{-6}~\mbox{M}_{\odot}~\mbox{yr}^{-1}.
\label{eq:M_f}
%\\
%& \approx &  7.2\times 10^{-6}~\mbox{M}_{\odot}~\mbox{yr}^{-1}.
%\nonumber
\ea
The numerical estimates are for $Q_0=1$. 
More general expressions for these variables for different 
opacity behavior ($\beta\neq 2$) and $\alpha_{\rm crit}$, as well as 
the coefficient in the expression (\ref{eq:alpha_GI}) for 
$\alpha_{\rm gt}$, can be found in Rafikov (2009).

\end{document}